\documentclass[doublecol]{epl2} 
\title{Electronic double-excitations in quantum wells: solving the two-time Kadanoff-Baym equations}
\shorttitle{Electronic double-excitations in quantum wells} %Insert here a short version of the title if it exceeds 70 characters

\author{K.~Balzer\inst{}, S.~Hermanns\inst{}, \and M.~Bonitz\inst{}}
\shortauthor{K.~Balzer~\etal}

\institute{
  \inst{}Institut f\"ur Theoretische Physik und Astrophysik - Christian-Albrechts-Universit\"at Kiel, Leibnizstrasse 15, 24098 Kiel, Germany
}

\usepackage{amssymb}
\usepackage{color}
\usepackage{graphicx}
\usepackage{booktabs}


\abstract{For a quantum many-body system, the direct population of states of double-excitation character is a clear indication that correlations importantly contribute to its nonequilibrium properties. We analyze such correlation-induced transitions by propagating the nonequilibrium Green's functions in real-time within the second Born approximation. As crucial benchmarks, we compute the absorption spectrum of few electrons confined in quantum wells of different width. Our results include the full two-time solution of the Kadanoff-Baym equations as well as of their time-diagonal limit and are compared to Hartree-Fock and exact diagonalization data.}

\newcommand{\nn}{\nonumber}
\newcommand{\mean}[1]{\langle{#1}\rangle}

\newcommand{\bra}[1]{\langle{#1}|}
\newcommand{\ket}[1]{|{#1}\rangle}

\newcommand{\intc}[1]{\int_{\cal C}\!\mathrm{d}{#1}\,}

\newcommand{\tr}[1]{\mathrm{Tr}[#1]}

\newcommand{\ii}{\mathrm{i}}

\renewcommand{\c}{{\cal C}}
\newcommand{\dc}{\delta_{\cal C}}
\newcommand{\tc}{T_{\cal C}}

\newcommand{\cG}{{\cal G}}

\begin{document}

\maketitle

\section{Introduction}
The prediction and interpretation of electronic excitations and photoemission (absorption) spectra~\cite{onida02} is becoming more and more vital for the study of (sub)femtosecond processes in atoms~\cite{krausz09}, (bio)molecules~\cite{kuleff11}, solids~\cite{wall11}, and nanoscale materials~\cite{fischer11}. Thereby, in many finite systems, particular importance must be placed on double-excitations (DEs) as they facilitate correlated excitation pathways which considerably enrich the nonequilibrium properties. The final state of such a transition is called a doubly-excited state, i.e., it is one which has dominant DE character but is not necessarily well described by a single doubly-excited Slater determinant, e.g.,~\cite{elliott11}. Prominent examples include the multiple electron-hole pair generation in semiconductors~\cite{klimov06} and the autoionization of atoms which involves the intermediate population of shake-up resonance states prior to fragmentation~\cite{bauch10}.

Aside from configuration interaction (CI) and exact diagonalization (ED) methods\footnote{CI and ED scale exponentially with system size and, hence, are unfavorable for the description of large systems.}~\cite{book_szaboostlund_mqc}, the calculation of electronic excitations is mainly based on time-dependent density functional theory (TDDFT), e.g.,~\cite{onida02,elliott11}, and many-body perturbation theory (MBPT) using Green's functions, e.g., \cite{onida02,strinati88} and references therein. In both approaches, the central quantity is the real-time, retarded two-particle response function~\cite{kwong00,saekkinen12} $\chi_{ij,kl}^{(2),\mathrm{R}}(t,t')={\delta\rho_{ij}^{(1)}(t)}/{\delta f_{lk}^{(1)}(t')}$, where $\rho^{(1)}$ denotes the one-particle reduced density matrix (1pRDM), $f^{(1)}$ is an external perturbation, and the poles of $\chi^{(2),\mathrm{R}}$ in frequency space refer to the excitation energies. The response function satisfies a Bethe-Salpeter equation (BSE)~\cite{strinati88} which contains a four-point integral kernel. For this reason, solutions are computationally challenging and often obtained at the expense of full frequency dependence and self-consistency, see ref.~\cite{onida02} for an overview. In TDDFT, the equations are simpler (being of two-point-type) but depend on the (generally unknown) exchange-correlation functional.

In this Letter, we apply nonequilibrium real-time Green's functions techniques to compute the excitation properties---in particular the DEs---of a correlated inhomogeneous quantum system in equilibrium. As shown in refs. \cite{bonitz99,kwong00}, this allows to avoid the direct solution of the BSE for $\chi^{(2),\mathrm{R}}$. Instead, we compute the first-order variation of the 1pRDM,
\begin{equation}
\label{ansatz}
\delta \rho_{ij}^{(1)}(t)=-\ii\delta\cG^{(1)}_{ij}(t,t^+)=-\ii\lim_{\epsilon\rightarrow 0}\delta\cG^{(1)}_{ij}(t,t+\epsilon)\;,
\end{equation}
in terms of the nonequilibrium Green's function (NEGF) $\cG^{(1)}$~\cite{kadanoffbaym62} following a suitable external excitation\footnote{The excitation spectrum is then obtained by inversion of the linear response relation.} $f^{(1)}(t)$. The key advantage of this method is that rather simple conserving approximations for the propagation of the NEGF (such as the second Born (2B) approximation) translate into high-level approximations for $\chi^{(2),\mathrm{R}}$ that obey the relevant sum rules \cite{kwong00}. This concept was applied to excitons in optically excited semiconductors \cite{kwong98}, the dynamic structure factor of the correlated electron gas~\cite{kwong00} and to the absorption spectrum of small atoms and molecules~\cite{dahlen06,dahlen07}.

A first extension of this real-time approach to DEs has been presented in ref.~\cite{saekkinen12} focusing on the spectral and response function of small Hubbard nanoclusters in the moderate-to-strong coupling regime. Yet no unambiguous criterion how to identify DEs in the spectrum has been given, and the accuracy and scope of validity of the used 2B for DEs has remained open. In this Letter, we answer these open questions by systematically analyzing electronic excitations in a few-particle quantum well (QW) structure as function of interaction strength (controlled by the QW width). We propagate the two-time NEGF fully including memory effects~\cite{stan09_jcp,balzer10_pra1, balzer10_pra2}. In order to achieve sufficiently long simulation times, we also consider the time-diagonal limit by employing the generalized Kadanoff-Baym ansatz (GKBA) \cite{lipavsky86}.

\section{Theory}
We start from a generic $N$-electron Hamiltonian $\hat{H}^{(N)}=\hat{H}^{(N)}_\alpha+\hat{H}^{(N)}_{\beta}$ composed of a non-interacting ($\alpha$) and an interacting part ($\beta$). The solution of the Schr\"odinger equation, $(\hat{H}^{(N)}-E_m^{(N)})\ket{\Psi^{(N)}_m}=0$, provides the complete spectral information including the ground (excited) state with energy $E_0^{(N)}$ ($E_{m>0}^{(N)}$). For a perturbation mediated by an operator $\hat{F}^{(N)}$, and for the system initially prepared in the state $\ket{\Psi^{(N)}_m}$, the full excitation spectrum, in the linear regime, consists of discrete peaks at frequencies $\omega_{m\rightarrow n}=\frac{1}{\hbar}|E_n^{(N)}-E_m^{(N)}|$ for which the transition moment $f_{m\rightarrow n}=\bra{\Psi_n^{(N)}}\hat{F}^{(N)}\ket{\Psi_m^{(N)}}$ is non-zero.

\subsection{Exact diagonalization}
For a given total spin $S$ and projection $M_S$, the wave function $\ket{\Psi^{(N)}_m}$ has the representation, $\ket{\Psi^{(N)}_m}=\hat{\Lambda}_-^{(N)}\{\ket{\psi^{(N)}_m}\ket{S,M_S}\}$, where $\hat{\Lambda}_-^{(N)}$ accounts for antisymmetrization and $\ket{\psi^{(N)}_m}$ describes the spatial part. If the Hamiltonian $\hat{H}^{(N)}$ is %being %exactly 
diagonalized in a product basis $\ket{P^{(N)}_{i_1\ldots i_N}}=\prod_{j=1}^{N}\ket{\phi_{i_j}^{(1)}}$ of one-particle states $\ket{\phi^{(1)}_i}$, the coefficients in the expansion $\ket{\Psi^{(N)}_m}=\sum_{i_1 \ldots i_N=1}^{M}c_{m,i_1\ldots i_N}^{(N)}\ket{P_{i_1\ldots i_N}^{(N)}}$ form eigenvectors of the $M^N\times M^N$ matrix $\bra{P^{(N)}_{i_1\ldots i_N}}\hat{H}^{(N)}\ket{P^{(N)}_{i_1'\ldots i_N'}}$. This is the procedure of the ED technique, where typically $M\gg N$ guarantees appropriate convergence. Further, as generally not all $M^N$ eigenvectors allow for the construction of a completely antisymmetric state, the relevant ground and excited state energies $E_m^{(N)}$ form an eigenvalue subset.

\subsection{Approximate excitation level (AEL)}
In many cases, the spectrum of the system can be classified according to the number of electrons that take part in the transition $m\rightarrow n$. Such a number is, e.g., provided by the AEL~\cite{stanton93},
\begin{eqnarray}
\label{eq.ael}
 A_{m\rightarrow n}&=&\frac{1}{2}\mathrm{Tr}\left|\rho^{(1)}_n-\rho^{(1)}_m\right|\;,
\end{eqnarray}
with\footnote{The prime indicates a second set of coordinates.} $\rho^{(1)}_{m}=\mathrm{Tr}_{2\ldots N}\ket{\psi^{(N)}_m}\bra{\psi^{(N)}_m}'$, where both 1pRDMs, $\rho^{(1)}_{m}$ and $\rho^{(1)}_{n}$, are expressed in the natural orbital basis that diagonalizes the initial state density matrix. While transitions with an AEL close to unity are referred to as single-excitations (SEs), an $A_{m\rightarrow n}\approx 2$ ($3$ or $4$ etc.) indicates states of essential double- (triple- or quadruple- etc.) excitation character. We note, that an integer AEL is only obtained in an effective single-particle approach using, e.g., the occupied and virtual Kohn-Sham or Hartree-Fock (HF) orbitals $\ket{\zeta_{i}^{(1)}}$ of the ground state. In this frozen-orbital (FO) or Koopmans' approximation~\cite{book_szaboostlund_mqc}, one or more electrons are promoted into virtual orbitals forming single singly- or multiply-excited Slater determinants. The corresponding ground (excited) state energy is approximate due to the neglect of orbital relaxations and correlations\footnote{Here, the prime implies summation over only those orbitals that contribute in the determinant.}:~$\tilde{E}_{m}^{(N)}=\sum_{i}' E^{(1)}_{\zeta,i}+\sum_{ij}'J_{ij}^{(2)}+\sum_{ij}'K_{ij}^{(2)}$, where $E_{\zeta,i}^{(1)}$ denote the orbital energies and $J_{ij}^{(2)}=\langle ii|\hat{h}_\beta^{(2)}|jj\rangle$ ($K_{ij}^{(2)}=\langle ij|\hat{h}_\beta^{(2)}|ji\rangle$) are the Coulomb (exchange) integrals with $\langle ij|\hat{h}_\beta^{(2)}|kl\rangle=h^{(2)}_{\beta,ijkl}=\int\!\!\!\int\!\mathrm{d}^3r\,\mathrm{d}^3r'\,\zeta_{i}(\vec{r})\,\zeta_{j}(\vec{r}\,')\,{h^{(2)}_\beta(\vec{r}-\vec{r}\,')}\,\zeta_{k}(\vec{r})\,\zeta_{l}(\vec{r}\,')$.

\subsection{Nonequilibrium Green's functions}
Since CI and ED can only handle small systems and pure states, in the following, we resort to a NEGF approach. As discussed in the introduction, eq.~(\ref{ansatz}) involves the NEGF $\cG^{(1)}_{ij}(t,t')=-\frac{\ii}{\hbar}\,\mean{\tc\,\hat{c}_i(t)\,\hat{c}^\dagger_j(t')}$ of the perturbed system, where $\hat{c}^{\dagger}_i$ ($\hat{c}_i$) denote fermionic creation (annihilation) operators, $i$ involves spin and orbital degrees of freedom and $\cG^{(1)}_{ij}(t,t')={\theta_\c(t-t')}\,{\cG^{(1),>}_{ij}(t,t')}-{\theta_\c(t'-t)}\,{\cG^{(1),<}_{ij}(t,t')}$. The NEGF is defined on the round-trip Keldysh contour $\c$~\cite{keldysh64} with the contour-ordering operator $\tc$ and obeys the two-time (non-Markovian) Kadanoff-Baym equations (KBEs)~\cite{kadanoffbaym62},
\begin{eqnarray}
\label{kbe}
 &&\{\ii\hbar\frac{\partial}{\partial t}\delta_{ik}-h^{(1)}_{\alpha,ik}(t)\}\,\cG^{(1)}_{kj }(t,t')\\
&=&\dc(t-t')\,\delta_{ij}+\intc{\bar{t}}\Sigma^{(1)}_{ik}(t,\bar{t})\,\cG^{(1)}_{kj}(\bar{t},t')\;,\nn
\end{eqnarray}
together with the adjoint equation with $t\leftrightarrow t'$. Here, $h^{(1)}_{\alpha,ij}(t)=\bra{i}\hat{h}_\alpha^{(1)}(t)\ket{j}$ denotes the one-particle energy and summation over $k$ is implied. A many-body approximation (MBA) for the two-time self-energy $\Sigma^{(1)}$ is obtained by MBPT as functional of the NEGF and the interaction potential $h^{(2)}_{\beta,ijkl}$. In HF approximation, one obtains $\Sigma^{(1),\mathrm{HF}}_{ij}(t,t')=\dc(t-t')\,\Sigma_{ij}^{(1),\mathrm{HF}}(t)$ with,
\begin{eqnarray}
\label{sigmahf}
\Sigma^{(1),\mathrm{HF}}_{ij}(t)=-\ii\hbar\,\{h^{(2)}_{\beta,ijkl}-h^{(2)}_{\beta,ilkj}\}\,\cG_{kl}^{(1)}(t,t)\;,\hspace{2pc}
\end{eqnarray}
whereas in the 2B approximation one adds,
\begin{eqnarray}
\label{sigma2b}
\Sigma^{(1),\mathrm{2B}}_{ij}(t,t')=-(\ii\hbar)^2\,h^{(2)}_{\beta,ikms}\,\{h^{(2)}_{\beta,ljrn}-h^{(2)}_{\beta,njrl}\}\times\nn\\
\times\;\cG^{(1)}_{kl}(t,t')\,\cG^{(1)}_{mn}(t,t')\,\cG^{(1)}_{rs}(t',t)\;.\hspace{1pc}
\end{eqnarray}
Both expressions conserve particle number, total momentum and energy. Computationally, the time-non-local, second-order self-energy (\ref{sigma2b}) involves a large number of ${\cal O}(n_\mathrm{b}^8)$ operations at basis dimension $n_\mathrm{b}$. However, its evaluation can be drastically simplified by using the finite element-discrete variable representation\footnote{Here, $h_{\beta,ijkl}^{(2)}$ becomes diagonal in, both, $ij$ and $kl$.} (FEDVR) that leads to a scaling behavior of ${\cal O}(n^4_\mathrm{b})$, see ref.~\cite{balzer10_pra1}. 

In this Letter, we apply two variants of solving the KBEs within approximations~(\ref{sigmahf}) plus (\ref{sigma2b}) for $\Sigma^{(1)}$. 
\begin{itemize}
 \item[I.] We, without any further approximations, propagate the NEGF in the double-time plane under the presence of the full memory kernel (r.h.s.~of the KBE). This gives rise to large memory requirements that limits the propagation time, though efficient code parallelization and FEDVR-type representations considerably extend the range of applicability, see ref.~\cite{balzer10_pra2}. To overcome this limitation, 
 \item[II.] we employ the GKBA\footnote{The GKBA retains the conservation properties \cite{bonitz-book}.}~\cite{lipavsky86}, $\cG^{(1),\gtrless}(t,t')\approx\pm\{\cG^{(1),\mathrm{R}}(t,t')\,\rho^{(1),\gtrless}(t')-\rho^{(1),\gtrless}(t)\,\cG^{(1),\mathrm{A}}(t,t')\}$, with $\rho^{(1),\gtrless}(t) = \pm\ii\hbar\,\cG^{(1),\gtrless}(t,t)$, where the Green's functions are reconstructed from their values on the time diagonal. This procedure is known to yield reliable results for weak to moderate coupling, e.g.,~\cite{kwong98,gartner06}. In the following, the retarded (advanced) function $\cG^{(1),\mathrm{R}}$ ($\cG^{(1),\mathrm{A}}$) is taken at the HF level allowing to  significantly reduce the computational effort in comparison to full 2B (case I) and allowing for longer propagation times combined with larger basis dimensions.
\end{itemize}
For details on the propagation of the NEGF in either cases, the reader is referred to~refs.~\cite{stan09_jcp, balzer10_pra2, hermanns12}.

To determine the excitation spectrum, we start the propagation at a time $t_0$, when the system is in the ground state\footnote{While, in case I, the initial state is obtained from the solution of the Dyson equation, see~\cite{balzer10_pra1}, for the GKBA (case II), we use adiabatic switching~\cite{rios11} for its generation.}. In the early stage of the evolution, the many-body system is perturbed in the form $h^{(1)}_{\alpha,ij}+f_0f^{(1)}_{ij}\delta(t-t_0)$. Complying with the constraint of sufficiently small amplitudes $f_0$, we then compute the linear response $\delta\rho^{(1)}_{ij}(t)=-\ii\{\cG^{(1),<}_{ij}(t,t)-\cG^{(1),<}_{ij}(t_0,t_0)\}$, and the Fourier transform of $\mean{\hat{f}^{(1)}}(t)=f_0\tr{\delta\rho^{(1)}(t) f^{(1)}}$ yields the absorption spectrum and the excited state energies.

\begin{table}
\caption{Four-electron QW with $\ket{S,M_S}=\ket{0,0}$ (singlet state) at coupling $\lambda^*=5$: Ground-state excitation energies $\omega_{0\rightarrow n}$ obtained from exact diagonalization (ED) and the frozen-orbital (FO) approximation. $A_{0\rightarrow n}$ denotes the AEL, eq.~(\ref{eq.ael}). DEs of dipole character are underlined. See also fig.~\ref{fig.1} (lower panel).}
\label{tab.1}
\begin{center}
\small
\begin{tabular}{ccccc}
\toprule
$n$ & $\omega^\mathrm{ED}_{0\rightarrow n}$ ($E_0^*$) & $A_{0\rightarrow n}$ & Character & $\omega^\mathrm{FO}_{0\rightarrow n}$ ($E_0^*$)\\
\midrule
 $1$ &  $1.1445$ & $1.079$ & dipole & $1.0616$\\
 $2$ &  $1.5478$ & $1.088$ & non-dipole & $1.5218$\\
 $3$ &  $2.1118$ & $1.842$ & non-dipole & $2.1491$\\
 $4$ &  $2.5559$ & $1.150$ & non-dipole & $2.4841$\\
 \underline{$5$} &  \underline{$2.6769$} & \underline{$1.870$} & \underline{dipole} & \underline{$2.5841$}\\
 $6$ &  $2.9794$ & $1.145$ & dipole & $2.9499$\\
 $7$ &  $3.0394$ & $1.969$ & non-dipole & $3.0646$\\
 \underline{$8$} &  \underline{$3.6314$} & \underline{$1.944$} & \underline{dipole} & \underline{$3.5629$}\\
 $9$ &  $3.9928$ & $1.997$ & non-dipole & $3.8556$\\
 $10$ & $4.2129$ & $1.956$ & non-dipole & $4.0034$\\
 $11$ & $4.3150$ & $1.119$ & dipole & $4.2735$\\
 \underline{$12$} & \underline{$4.5544$} & \underline{$2.048$} & \underline{dipole} & \underline{$4.4921$}\\
 $13$ & $4.7619$ & $1.122$ & non-dipole & $4.7512$\\
 $14$ & $4.9300$ & $1.998$ & non-dipole & $4.9824$\\
 $15$ & $4.9575$ & $2.830$ & dipole & $4.9350$\\
 $16$ & $5.4099$ & $2.797$ & non-dipole & $5.3985$\\
 $17$ & $5.4250$ & $2.034$ & non-dipole & $5.3527$\\
 \underline{$18$} & \underline{$5.5413$} & \underline{$2.019$} & \underline{dipole} & \underline{$5.4311$}\\
 \underline{$19$} & \underline{$5.8032$} & \underline{$2.038$} & \underline{dipole} & \underline{$5.6819$}\\
 $20$ & $5.8882$ & $2.153$ & non-dipole & $5.7953$\\
 \underline{$21$} & \underline{$6.0119$} & \underline{$1.981$} & \underline{dipole} & \underline{$5.9254$}\\
 $22$ & $6.3480$ & $2.055$ & non-dipole & $6.2934$\\
 $23$ & $6.4185$ & $2.839$ & non-dipole & $6.3540$\\
 $24$ & $6.4711$ & $1.096$ & non-dipole & $6.4505$\\
 $25$ & $6.7825$ & $2.802$ & non-dipole & $6.7618$\\
 \underline{$26$} & \underline{$6.8345$} & \underline{$2.018$} & \underline{dipole} & \underline{$6.7711$}\\
 $27$ & $6.8966$ & $2.941$ & dipole & $6.8230$\\
 $28$ & $6.9333$ & $1.104$ & dipole & $6.9334$\\
 $29$ & $7.1216$ & $2.108$ & non-dipole & $7.0143$\\
 $30$ & $7.2410$ & $2.912$ & dipole & $7.2248$\\
 $31$ & $7.4516$ & $2.035$ & non-dipole & $7.2317$\\
 \underline{$32$} & \underline{$7.5683$} & \underline{$1.986$} & \underline{dipole} & \underline{$7.5295$}\\
 $33$ & $7.7324$ & $3.876$ & non-dipole & $7.7465$\\
 \underline{$34$} & \underline{$7.7932$} & \underline{$2.150$} & \underline{dipole} & \underline{$7.7255$}\\
 $35$ & $7.9891$ & $2.031$ & non-dipole & $7.8790$\\
\bottomrule
\end{tabular}
\end{center}
\label{tab.1}
\end{table}

\begin{figure*}[t]
\includegraphics[width=0.985\textwidth]{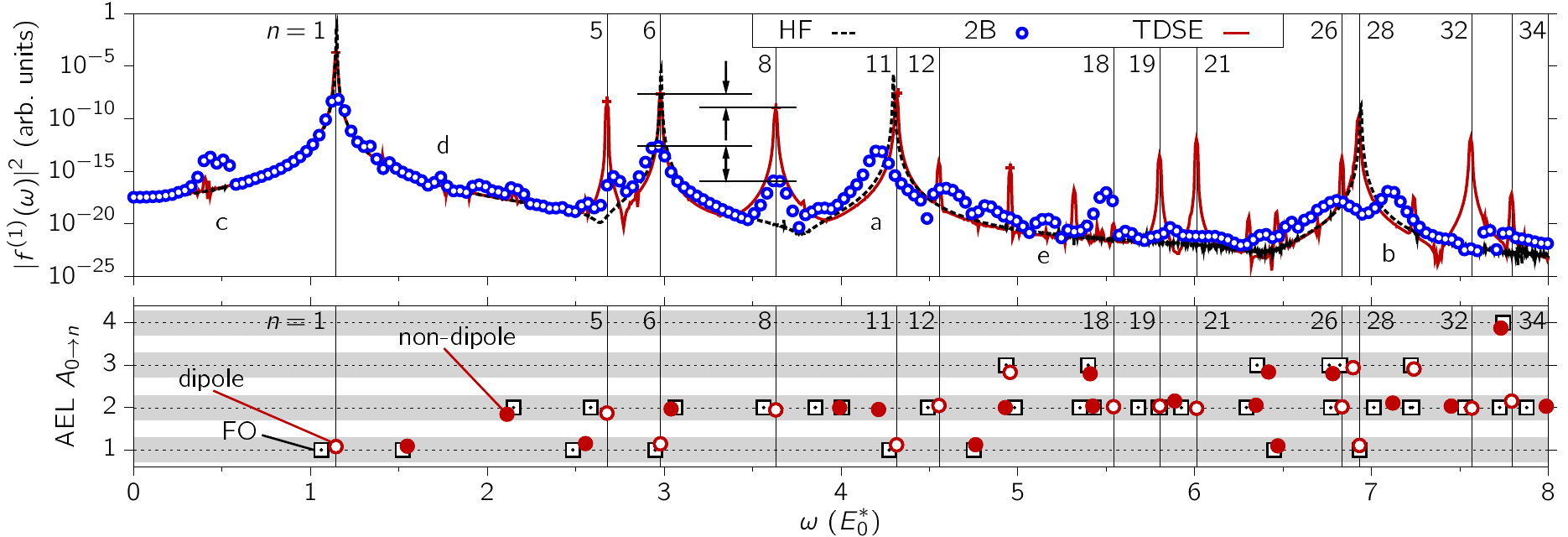}
\vspace{-0.65pc}
\caption{(Color online) Ground-state excitation spectrum of the four-electron QW at $\lambda^*=5$. \underline{Lower panel}: approximate excitation level, eq.~(\ref{eq.ael}). The red open (closed) dots indicate states that are accessible from the ground state by (non-)dipole transitions. The black squares show the excitation energies in FO approximation (integer AEL). All dipole-allowed SEs and DEs (cf.~the black vertical lines) are numbered according to table~\ref{tab.1}. \underline{Upper panel}: Dipole excitation spectrum $|f^{(1)}(\omega)|^2$ as obtained from the time-dependent Schr\"odinger equation (TDSE) and from the Kadanoff-Baym equations in HF (black dashed line) and 2B approximation (blue dots). The labels {\sffamily a}-{\sffamily e} mark prominent deviations of the 2B result from HF and TDSE, respectively.}
\label{fig.1}
\end{figure*}

\section{Model}
As test system, we consider $N=4$ electrons (charge $e$, effective mass $m^*$, positions $\vec{r}_i$) with singlet spin configuration $\ket{S,M_S}=\ket{0,0}$ in a quantum well (QW) potential of width $L$. Neglecting the lateral electron motion, the Hamiltonian in units of the confinement energy\footnote{$E_0^*\pi^2/2$ is the ground-state energy of a single-electron QW.} $E_0^*=\hbar^2/(m^*L^2)$ reads in vertical direction~$\hat{H}^{(N)}(t)=\sum_{i=1}^{N}\hat{h}^{(1)}_\alpha(z_i,t)+\lambda^*\sum_{1\leq i<j}^{N}\hat{h}^{(2)}_\beta(z_i-z_j)$, with $\vec{r}_i=(0,0,z_iL)$, $\hat{h}^{(1)}_\alpha(z,t)=-\frac{1}{2}\frac{\partial^2}{\partial z^2}+f_0 f^{(1)}(z)\,\delta(t-t_0)$ and ${\hat{h}^{(2)}_\beta(z-z')}=[(z-z')^2+\kappa^2]^{-1/2}$. The coupling strength $\lambda^*=L/a_0^*=e^2m^*L/(4\pi\varepsilon_0\varepsilon^*\hbar^2)$ with effective Bohr radius $a_0^*$ and (material) dielectric constant ($\varepsilon^*$) $\varepsilon_0$ defines the relative interaction strength between the electrons. While $\lambda^*\rightarrow0$ represents the ideal quantum regime, the limit $\lambda^*\rightarrow\infty$ leads to quasi-classical, Wigner-crystal behavior. In a GaAs heterostructure ($m^*=0.067m_\mathrm{e}$, $\varepsilon^*=12.9$), the moderate-to-strong coupling cases $\lambda^*=1$ ($5$) correspond to well widths of ${L={10.2\,\textup{nm}}}$ (${50.9\,\textup{nm}}$) which are realistic QW dimensions at confinement energies ${E_0^*={10.956\,\textup{meV}}}$ (${0.438\,\textup{meV}}$). Further, the soft-Coulomb form of $h^{(2)}_\beta$ with regularization $\kappa$ (we set $\kappa=0.2$) prevents the divergence of the two-electron integrals which are constructed from one-dimensional FEDVR spin-orbitals.

In the following, we consider $f^{(1)}$ as the dipole operator, i.e., $f^{(1)}(z)=z$ such that $f_0$ is measured in units of $eL{\cal E}_0/E_0^{*}$, where ${\cal E}_0$ is the electric field strength. The corresponding transition dipole moment (TDM) is defined by $f^{(N)}_{m\rightarrow n}=e\bra{\psi_n^{(N)}}\sum_{i=1}^{N}z_i\ket{\psi_m^{(N)}}$. For transitions of dipole (non-dipole) type, it is $|f^{(N)}_{m\rightarrow n}|^2>0$ ($|f^{(N)}_{m\rightarrow n}|^2\equiv0$).

\section{Results}
First, we concentrate on the moderate-to-strong coupling case $\lambda^*=5$ which is typical for sub-$100\,\textup{nm}$ semiconductor heterostructures. In contrast to the ideal, non-interacting system\footnote{The ideal ground-state energy is $\pi^2/5\,E_0^*$.}, where dipole selection rules prohibit DEs, we, here, expect doubly-excited states of relevant TDM. Table~\ref{tab.1} shows all ground-state excitation energies $\omega^\mathrm{ ED}_{0\rightarrow n}$ below $8E_0^*$ as obtained from ED itemized by AEL and general dipole character. We observe, that dipole and non-dipole transitions alternate and that the AEL allows for a unique classification of each excited state. Overall, its deviation from an integer is less than $0.21$.

According to table~\ref{tab.1}, the dipole excitation spectrum starts with a SE {($n=1$)} of energy $\omega^\mathrm{ED}_{0\rightarrow1}=1.14\,E_0^*$ followed by a DE ($n=5$) with $\omega^\mathrm{ED}_{0\rightarrow5}=2.68\,E_0^*$, a SE ($n=6$) with $\omega^\mathrm{ED}_{n\rightarrow6}=2.80\,E_0^*$, and a DE ($n=8$) with $\omega^\mathrm{ED}_{0\rightarrow8}=3.63\,E_0^*$, etc.. The first dipole-allowed triple-excitation has quantum number $n=15$. Further, we note, that the energetically lowest doubly-{(quadruply-)}excited state with $n=3$ ($33$) is not of dipole-type, cf.~also fig.~\ref{fig.1} (lower panel). Moreover, table~\ref{tab.1} includes the result of the FO approximation as carried out using the ground-state HF orbitals, see $\omega^\mathrm{FO}_{0\rightarrow n}$. Thereby, the energies well approach the exact ones although FO can lead to wrong ordering, cf.~$n=14$ vs.~$15$ ($16$ vs.~$17$), and can produce shifts as large as $0.2\,E_0^*$, cf.~$n=10$ and $31$. 

The situation for $\lambda^*=5$ is visualized in the lower and upper panel of fig.~\ref{fig.1} including exact and the approximate NEGF results. From the linear-response dipole spectrum as obtained from the solution of the time-dependent Schr\"odinger equation (TDSE), see the red solid line for $|f^{(1)}(\omega)|^2$ (upper panel), we find that DEs appear indeed with significant TDM, cf., e.g., the peaks with $n=5$ and $8$. From the solution of the KBEs~(\ref{kbe}) according to case~I, we obtain the time-dependent HF and 2B result, which, respectively, are indicated by the black dashes and the blue dots. First, we observe that HF has only the ability to describe SEs, cf.~the peaks for states with $n=1$, $6$, $11$ and $28$. Thereby, the corresponding energies are more accurate than those in FO approximation due to the inclusion of orbital relaxations. The failure of reproducing peaks of multiply-excited states, however, leads overall to a very simple spectrum. In contrast, the 2B approximation, which accounts for more than $70$\% of the correlation energy of the QW's electronic ground state\footnote{At $\lambda^*=5$, the exact ground-state energy is $5.5278\,E_0^*$. The approximate treatment yields in HF:~$5.5384\,E_0^*$, in 2B (case~I):~$5.5306\,E_0^*$, and in GKBA (case~II):~$5.5298\,E_0^*$.}, shows much more structure in the spectrum. In particular, it yields additional peaks that are located at energies for which we expect DEs, cf.~the states indicated $n=5$, $8$ and $12$.

\begin{figure}[t]
\hspace{1.0pc}\includegraphics[width=0.435\textwidth]{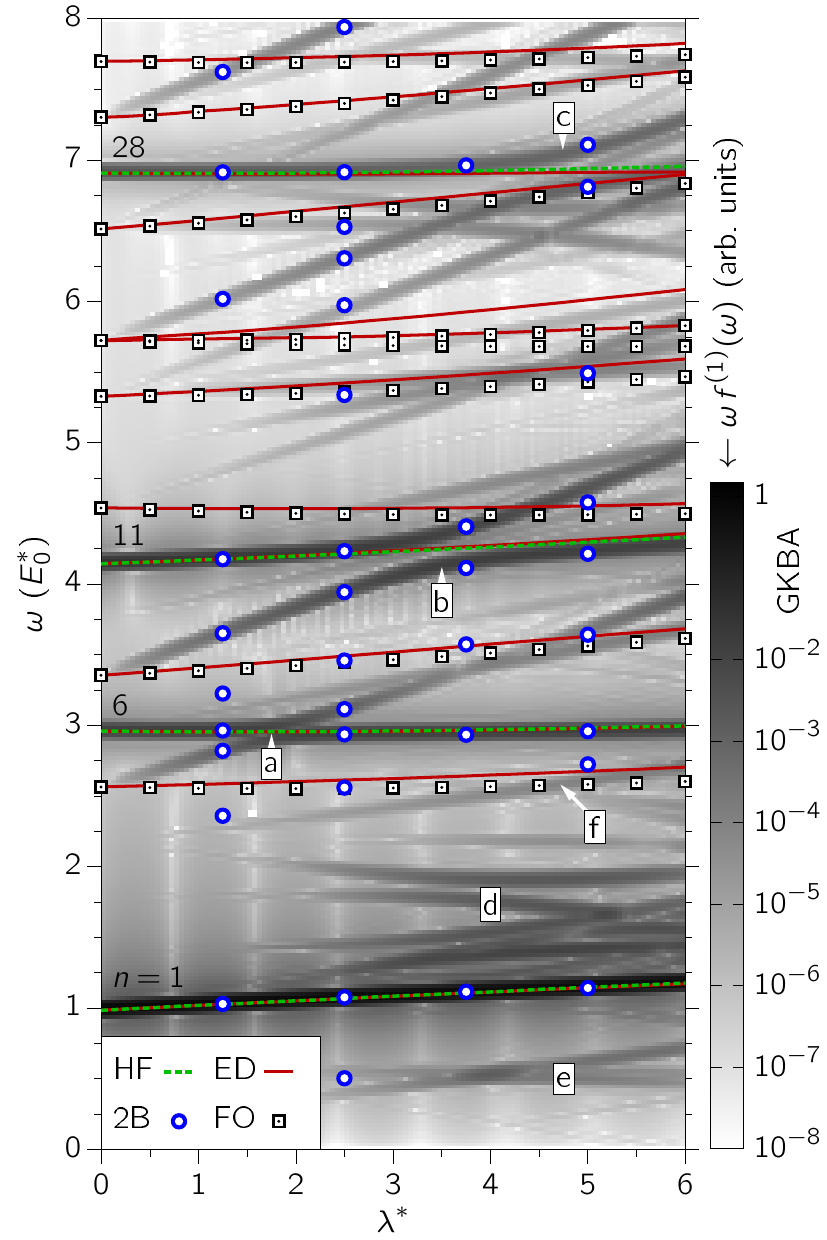}
\vspace{-0.65pc}
\caption{(Color online) Ground-state dipole excitation spectrum of the four-electron QW as function of the coupling strength $\lambda^*$. HF:~green dashed lines, GKBA (case~II):~gray-scale density plot, full 2B (case~I):~blue dots, ED:~red solid lines, and FO for DEs  with $A_{0\rightarrow n}=2$:~black squares. Note, that, for ED, only SEs and DEs are shown. While labels {\sffamily a}-{\sffamily c} mark the avoided crossings in the GKBA result, labels {\sffamily d}-{\sffamily f} indicate excitations which are not attributable to DEs.}
\label{fig.2}
\end{figure}

It is tempting to identify the additional peaks in 2B as the correct DEs. However, great care is needed---especially in the context of additional deficiencies in the 2B result which require explanations:~(i) the relatively low excitation strength for the peaks at $n=5$ and $8$ relative to those at $n=6$ and $11$ (compare with the TDSE data), (ii) the observation of energy shifts and splittings for singly-excited states which are well covered even by the HF result (see labels {\sffamily a}, {\sffamily b}), and (iii) the presence of peaks at low and high energies which cannot be attributed to DEs of the system (cf.~labels {\sffamily c}, {\sffamily d} and {\sffamily e}).

With the intention to explain these shortcomings, we systematically investigate the excitation spectrum as function of $\lambda^*$. Focusing on SEs and DEs, fig.~\ref{fig.2} shows how the spectrum changes when going from the non-interacting system to the moderate-to-strong coupling case $\lambda^*=6$. In HF approximation, the SEs with $n=1$, $6$, $11$ and $28$ are very well described over the whole $\lambda^*$-range, see the green dashed lines which practically lie on top of the exact result (red lines). For small $\lambda^*\lesssim1$, the same is found for the GKBA result (gray-scale density) such that, in the limit $\lambda^*\rightarrow0$, for, both, HF and GKBA, the SE energies exponentially converge towards the ones of the ideal system. For doubly-excited states, we find a completely different behavior. Comparing GKBA to ED and to FO (black squares), we observe that the DE energies as described by GKBA, vary too strongly with the coupling strength though the correct $\lambda^*=0$-limit is reached. Even ''avoided'' crossings of the $N$-particle energy levels occur for $\lambda^*\approx1.75$ and thereafter for $\lambda^*\approx3.5$ and $5$, cf.~labels {\sffamily a}, {\sffamily b} and {\sffamily c}. Such a behavior is not supported by the exact description\footnote{At least in the considered regime of coupling strengths.}. However, it is not a deficiency of the GKBA because full 2B calculations (see the blue dots which mark resolvable peak positions) confirm this behavior.

\begin{figure}[t]
\includegraphics[width=0.475\textwidth]{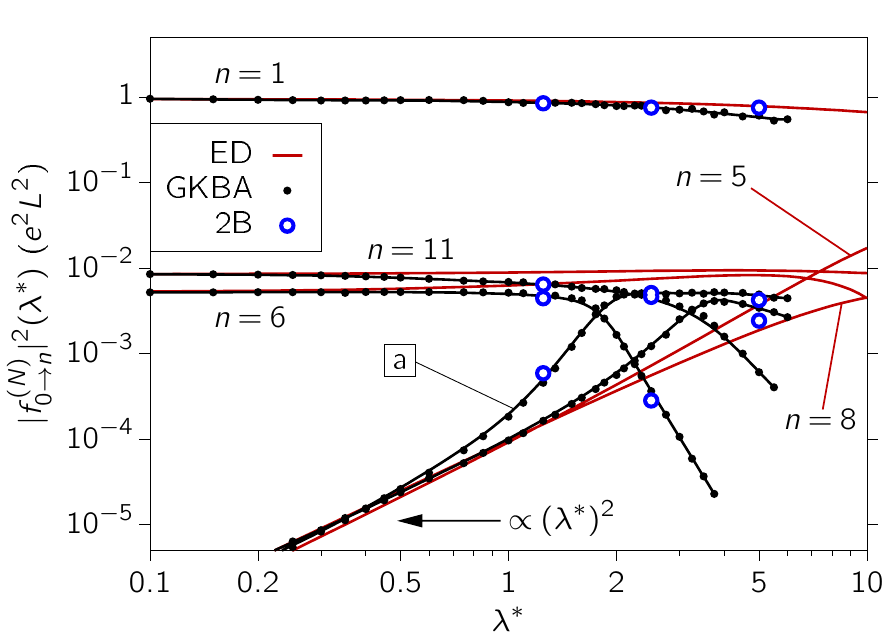}
\vspace{-0.65pc}
\caption{(Color online) Four-electron QW as in figs.~\ref{fig.1} and~\ref{fig.2}. Shown is the absolute square of the TDMs for the five energetically lowest excited states as function of the coupling strength $\lambda^*$; states with $n=1,6,11$:~SEs, $n=5,8$:~DEs. GKBA:~black lines, 2B:~blue dots, and ED:~red solid lines. For $\lambda^*>2$, the TDM for the energetically lowest DE (label {\sffamily a}) indicates a transition from a DE to a SE, compare with fig.~\ref{fig.2}.
}
\label{fig.3}
\end{figure}

To prove that the non-singly excited states found by GKBA and 2B are really of DE character, we, in fig.~\ref{fig.3}, analyze the TDMs of the five lowest excited states as function of $\lambda^*$. First, for the three SEs below $\lambda^*\approx1$, the exact TDMs are nearly constant, see the red lines indicated $n=1$, $6$ and $11$. Second, for the two DEs ($n=5$ and $8$), $|f^{(N)}_{0\rightarrow n}|^2$ is several orders of magnitude smaller for low coupling strength and increases with $(\lambda^*)^2$. Both these properties can be understood by standard perturbation theory for the four-electron QW and are well reproduced by the GKBA calculations, see the black lines. Beyond a coupling strength of $\lambda^*\approx0.5$, however, the approximate and exact TDMs start to deviate considerably. Here, GKBA shows a complex $\lambda^*$-dependence whereby the energetically lowest (isolated) excited state with quantum number $n=1$ is less affected. We attribute this to the ''avoided'' energy levels crossings, where the energetically lowest DE takes over the role of the second SE (cf.~curve {\sffamily a} and compare with fig.~\ref{fig.2}). The same occurs for the second DE at $\lambda^*\approx4$.

\section{Discussion}
Following figs.~\ref{fig.2} and~\ref{fig.3}, we can clearly distinguish between states of different excitation character in the GKBA (2B) spectrum and conclude that reasonable results are obtained for small coupling. Moreover, we are now in a position to give reasons for some of the deficiencies formulated for the $\lambda^*=5$-case. Obviously, the presence of an ''avoided'' crossing for the third (fourth) SE in the case of 2B and GKBA as function of $\lambda^*$ explains the energy shift (splitting) in the corresponding excitation energy, cf.~label {\sffamily a} ({\sffamily b}) in fig.~\ref{fig.1}. The fact that the 2B approximation underestimates the TDMs for several excited states is of similar reason, see, e.g., the GKBA result for the regime $\lambda^*>2$ in fig.~\ref{fig.3}. A further main results is that, due to the strong coupling dependence of the DE energies in 2B, the peak at $\omega\approx3.6\,E_0^*$ ($n=8$) in fig.~\ref{fig.1} (blue dots) does not originate from the second but from the first DE, compare with fig.~\ref{fig.2}. In this regard, we also note that practically all energy differences $\omega_{0\rightarrow n}^\mathrm{2B}-\omega_{0\rightarrow n}^\mathrm{ED}$ for doubly-excited states do not exponentially become small in the limit $\lambda^*\rightarrow0$, although this is the case for all SEs. Instead, it seems that DEs have in 2B an intrinsic coupling dependence determined by perturbation character of the MBA. Furthermore, we mention, that the additional peaks emerging in the 2B result disappear for small $\lambda^*$ but cannot be attributed to states of the real system\footnote{Note, that these artifacts can mimic relevant excited states, cf., e.g., label {\sffamily f} in fig.~\ref{fig.2} ($\lambda^*=5$) which refers to the first DE.}, cf.~labels {\sffamily c}-{\sffamily e} in fig.~\ref{fig.1} (upper panel) and labels {\sffamily d} and {\sffamily e} in fig.~\ref{fig.2}. The study of their origin requires further investigations\footnote{Similar effects have been observed in the 2B treatment of excitations in small Hubbard chains, cf.~ref.~\cite{hermanns12}.}---however, we report that their presence does not depend on the approximation level of the ground state at $t=t_0$, i.e., whether it is of HF-type or correlated.

In conclusion, the four-electron QW is sufficiently simple to allow for ED results. This has given us the possibility to quantify the accuracy of the 2B approximation with respect to important correlation features such as electronic DEs, which gain importance for moderate to strong coupling (cf.~fig.~\ref{fig.3}). The spectrum follows from direct time-propagation of the NEGF, where the GKBA has allowed for a detailed analysis for a broad range of interaction strengths (tuned by the QW width). DEs have been identified by tracing the system behavior to the zero-coupling limit analyzing the functional dependence of their oscillator strengths. In contrast to ED, we, in the 2B result (with and without the GKBA), have found a strong coupling dependence of the DE energies and, as a consequence, mixings (or hybridizations) of SEs and DEs which show up as ''avoided'' crossings, cf.~fig.~\ref{fig.2}. As this occurs in the context of well-described SEs in a HF treatment, great care is needed when interpreting correlated excitation pathways in MBAs beyond HF. It will be interesting to see, whether and how much the present behavior can be ''repaired'' by more advanced many-body self-energies such as the T-matrix or $GW$ approximation.

\acknowledgments
This work was supported by the Deutsche Forschungsgemeinschaft via grant BO 1366-9 and by computing time at the North-German Supercomputing Alliance (HLRN) via Grant No.~shp0006.

\end{document}